\documentclass[12pt,preprint]{aastex}

\newcommand{\cm}{\,{\rm cm}^{-3} } 
\newcommand{\msun}{\thinspace M_\odot} 
\newcommand{\km  }{\,{\rm km\, s^{-1}} } 
  
\newcommand{\etal }{{et al.} }

\shorttitle{Low mass stars}
\shortauthors{Machida 2010}

\begin{document}
\title{Recent Developments in Simulations of Low-mass Star Formation}
\author{Masahiro N. Machida\altaffilmark{1}} 
\altaffiltext{1}{National Astronomical Observatory of Japan, Mitaka, Tokyo 181-8588, Japan; masahiro.machida@nao.ac.jp}

%%\title[Low mass stars] %% give here short title %%
%%{Recent Developments in Simulations of Low-mass Star Formation}
%%
%%\author[Masahiro N. Machida]   %% give here short author list %%
%%{Masahiro N. Machida$^1$
%%  \thanks{Present address: Fluid Mech Inc., 24 The Street, Lagos, Nigeria.},
%%}
%%\affiliation{$^1$
%%National Astronomical Observatory of Japan, Mitaka, Tokyo 181-8588, Japan \\ email; {\tt masahiro.machida@nao.ac.jp}
%%}
%%\pubyear{2010}
%%\volume{xxx}  %% insert here IAU Symposium No.
%%\pagerange{}
% \date{?? and in revised form ??}
%%\setcounter{page}{1}
%%\jname{Title of your IAU Symposium}
%%\editors{A.C. Editor, B.D. Editor \& C.E. Editor, eds.}
%%\begin{document}
%%\maketitle

\small
\begin{abstract}
In star forming regions, we can observe different evolutionary stages of various objects and phenomena such as molecular clouds, protostellar jets and outflows, circumstellar disks, and protostars. 
However, it is difficult to directly observe the star formation process itself, because it is veiled by the dense infalling envelope. 
Numerical simulations can unveil the star formation process in the collapsing gas cloud. 
Recently, some studies showed protostar formation from the prestellar core stage, in which both molecular clouds and protostars are resolved with sufficient spatial resolution. 
These simulations showed fragmentation and binary formation, outflow and jet driving, and circumstellar disk formation in the collapsing gas clouds. 
In addition, the angular momentum transfer and dissipation process of the magnetic field in the star formation process were investigated. 
In this paper, I briefly review recent developments in numerical simulations of low-mass star formation.
\end{abstract}


%%\firstsection % if your document starts with a section,
%%              % remove some space above using this command.

\section{Introduction}
A star is born in a molecular cloud core through gravitational contraction.
Molecular clouds that are the initial state of the star formation are frequently observed in various star-forming regions, and we have much information about them.
In star-forming regions, we have also observed various objects and phenomena, such as protostars, protostellar jets, bipolar outflows, and circumstellar disks.
%% resulting from the gravitational contraction of molecular cloud cores.
They are the outcome of the gravitational contraction of molecular cloud cores.
Thus,  numerous observational studies have allowed us to understand both the initial state and its outcome for star formation.
However, observations do not allow us to understand the star-formation process itself (or gravitational contraction phase), because (proto)star formation occurs in dense cloud cores, which are difficult to observe directly.
%% thus necessitating a theoretical approach.
Thus, theoretical approach is necessary to understand the star formation process.
In order to understand the star-formation process in a collapsing cloud core, in addition to the self-gravity of the contracting gas, we have to consider the effects of thermal pressure, the Lorentz force, and rotation, which are all intricately interrelated.
%% to understand the star formation in the collapsing cloud core.
%%In the star formation process, these processes are intricately interrelated each other.
Therefore, we need detailed numerical simulations to unveil the star-formation process in a dense collapsing cloud core.

However, it is difficult to calculate the star-formation process from the molecular cloud core (the prestellar core stage) to protostar formation (the protostellar phase) through the runaway gravitational collapse, because we have to resolve the spatial scale over $\sim$7 orders of magnitude and the density scale over $\sim$18 orders of magnitude.
%%This is because these numerical simulations are so difficult is because they 
Molecular clouds have sizes of $\sim10^{5}$AU and densities of $\sim10^{4}\cm$, while protostars have sizes of $\sim0.01$\,AU and densities of $\sim10^{22}\cm$.
Thus, we require special numerical techniques, such as AMR (Adaptive Mesh Refinement) and SPH (Smoothed Particle Hydrodynamics), to spatially resolve both the molecular cloud core and the protostar.
Now, using these methods, we can unveil the star-formation process by directly calculating the star formation starting from the prestellar core stage.
This review summarizes recent developments in numerical simulations of low-mass star formation in the collapsing cloud cores.

\section{Outline of Protostar Formation}
\label{sec:outline}
At first, based on the results of spherically symmetric calculations \citep{larson69,masunaga00}, I briefly outline the low-mass star formation process and its thermal evolution (for details, see Fig.~2 of \citealt{masunaga00} along with its explanation).
Observations indicate that stars are born in molecular cloud cores that have number densities of $n\sim10^4\cm$.
After the gravitational collapse begins, the density of the cloud increases with time.
In the cloud core, the gas collapses isothermally and remains at $\sim10$\,K until the number density reaches $n\sim10^{10}\cm$; at this point, the central region becomes optically thick and the equation of state becomes hard.
Then, the first adiabatic core (the so-called first core) with a size of $\sim1$\,AU appears.
Subsequent to the formation of the first core, further rapid collapse is induced in a small central part of the first core because of the dissociation of molecular hydrogen when the number density exceeds $n\gtrsim10^{16}\cm$.
Finally, the gas becomes adiabatic again for $n\gtrsim10^{21}\cm$ because of the complete of the dissociation of molecular hydrogen, and a protostar with a size of $\sim0.01$\,AU appears in the collapsing cloud core.

At its formation, the first core has a mass of $\sim0.1-0.01\msun$, while the protostar has a mass of $\sim10^{-3}\msun$, which corresponds to the Jovian mass.
Thus, the massive first core ($\sim 1-10$\,AU) encloses the protostar ($\sim0.01$\,AU).
A spherically symmetric calculation, which could not include the effect of the rotation, showed that the first core gradually shrinks and disappears in $\sim10$\,yr after the protostar formation.
On the other hand, multidimensional calculations, which did include the effect of the rotation, showed that the first core remains long after the protostar formation (\citealt{saigo06}), evolving into the circumstellar disk in the main accretion phase (\citealt{machida10}).
In the main accretion phase, the protostar acquires almost all its mass by gas accretion, reaching $\sim1\msun$.
In the subsequent sections, I review recent developments in low-mass star-formation simulations, especially in the gas-collapsing phase.

\section{Fragmentation and Binary Formation}
Observations have shown that the multiplicity of pre-main sequence stars is larger than that of main-sequence stars in  star-forming regions (e.g., \citealt{mathiu94})
Recently, extremely young protostars (i.e., Class 0 protostars) have been observed with radio interferometers (e.g., \citealt{looney00}) and wide-field near-infrared cameras (\citealt{duchene04}).
These observations showed that stars already have a high multiplicity at the moment of their birth.
Thus, we expected that a large fraction of stars are born as binary or multiple systems.

\begin{figure}[b]
%%\vspace*{-0.5 cm}
\begin{center}
 \includegraphics[width=150mm]{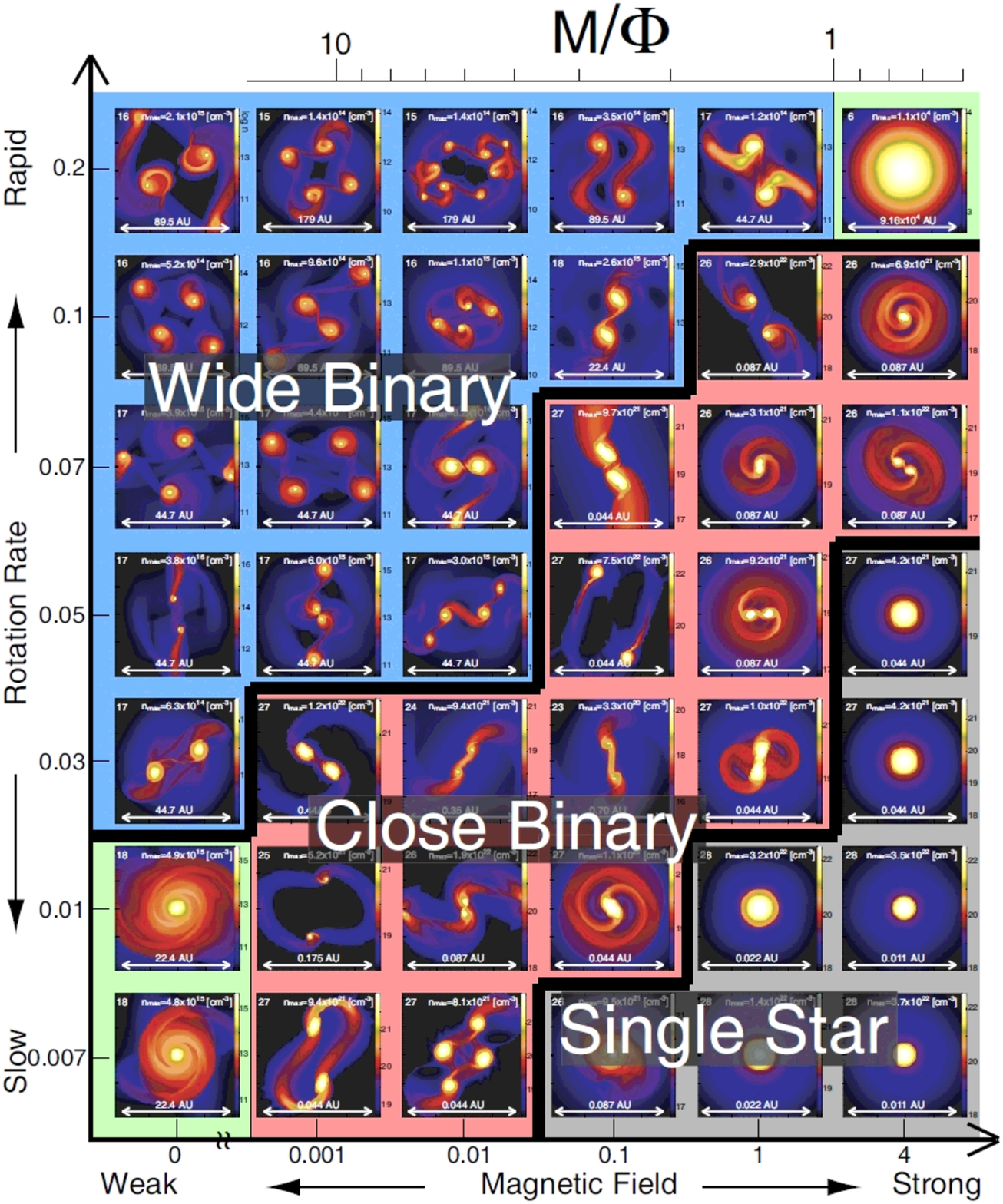} 
%%\vspace*{-0.1 cm}
\caption{
Final states for clouds with initially different magnetic fields ($x$-axis) and angular velocities ($y$-axis).
The densities (color-scale) on the cross section of the $z=0$ plane are plotted in each panel.
Background colors indicate the following: fragmentation occurs with separation of $>1$\,AU, resulting in wide binaries (blue); fragmentation occurs with separation of $<1$\,AU, resulting in close binaries (pink);  no fragmentation occurs through all phases of the cloud evolution, resulting in single-stars (gray); and the cloud no longer collapses, resulting in no star formation (green).
The upper horizontal axis indicates the initial mass-to-flux ratio that is normalized by the critical value.
}
\label{fig:1}
\end{center}
\end{figure}

It is considered that rotation causes fragmentation in a collapsing cloud, which then lead to the formation of binary or multiple star systems.
%%binary or multiple system is formed through fragmentation in the collapsing cloud with rotation.
Several three-dimensional simulations of the evolution of rotating collapsing clouds have investigated the possibility of fragmentation and binary formation (see, review of \citealt{bodenheimer00} and \citealt{goodwin07}).
\citet{miyama84} and \citet{tsuribe99} calculated the evolution of spherical clouds in the isothermal regime with initially uniform density and rigid-body rotation.
They found that fragmentation (and thus binary formation) occurs in the isothermal contracting phase ($n<10^{10}\cm$) only when the initial cloud is (highly) thermally unstable against gravity (see, also \citealt{boss93}).
However, fragmentation easily occurs after the gas becomes adiabatic ($n>10^{10}\cm$, e.g., \citealt{matsu03}).
Several studies have shown that, without the magnetic field, fragmentation frequently occurs in the adiabatic phase even when the molecular cloud has a small angular momentum.
This is because after the gas becomes adiabatic, it collapses very slowly and the perturbation that  induces fragmentation can grow.
In addition, in the adiabatic phase, the cloud rotation can form a disk sufficiently thin for fragmentation to occur.
On the other hand, recent magnetohydrodynamics simulations have shown that the magnetic field suppresses fragmentation and binary formation (\citealt{hosking04}, \citealt{machida04}, \citealt{machida05b}, \citealt{machida08a}, \citealt{hennebelle08b} and \citealt{price07}).
This is because the angular momentum that could lead to the formation of a disk thin enough for fragmentation to occur is transferred by magnetic braking and the protostellar outflow (see, \S\ref{sec:outflow}).
Thus, in a strongly magnetized cloud, no thin disk appears (\citealt{mellon09}) and, therefore, no fragmentation occurs.
Figure~\ref{fig:1} shows the rotation and magnetic field conditions under which fragmentation occurs with the different panels showing the final state of clouds with initially different rotational and magnetic energies.
This figure indicates that a large cloud rotation rate promotes fragmentation, but a strong cloud magnetic field suppresses it.
In other words, a molecular cloud with a strong magnetic field must have a large angular momentum in order to form binary systems.

\section{Protostellar Outflows and Jets}
\label{sec:outflow}
The observations indicate that protostellar outflows are ubiquitous in star-forming regions.
Flows originating from protostars are typically classified into two types: molecular outflows observed mainly through line emission from their CO molecules (\citealt{arce06}), and optical jets observed through their optical emission (\citealt{pudritz06}).
Molecular outflows exhibit wide opening angles and slow velocities  ($10-50\km$, e.g., \citealt{belloche02}), while optical jets exhibit good collimation and high velocities  ($100-500\km$, e.g., \citealt{bally07}).
The observations also indicate that around each protostar, a wide-opening-angle low-velocity outflow encloses a narrow-opening-angle high-speed jets (\citealt{mundt83}).

\begin{figure}[b]
%%\vspace*{-0.3 cm}
\begin{center}
 \includegraphics[width=160mm]{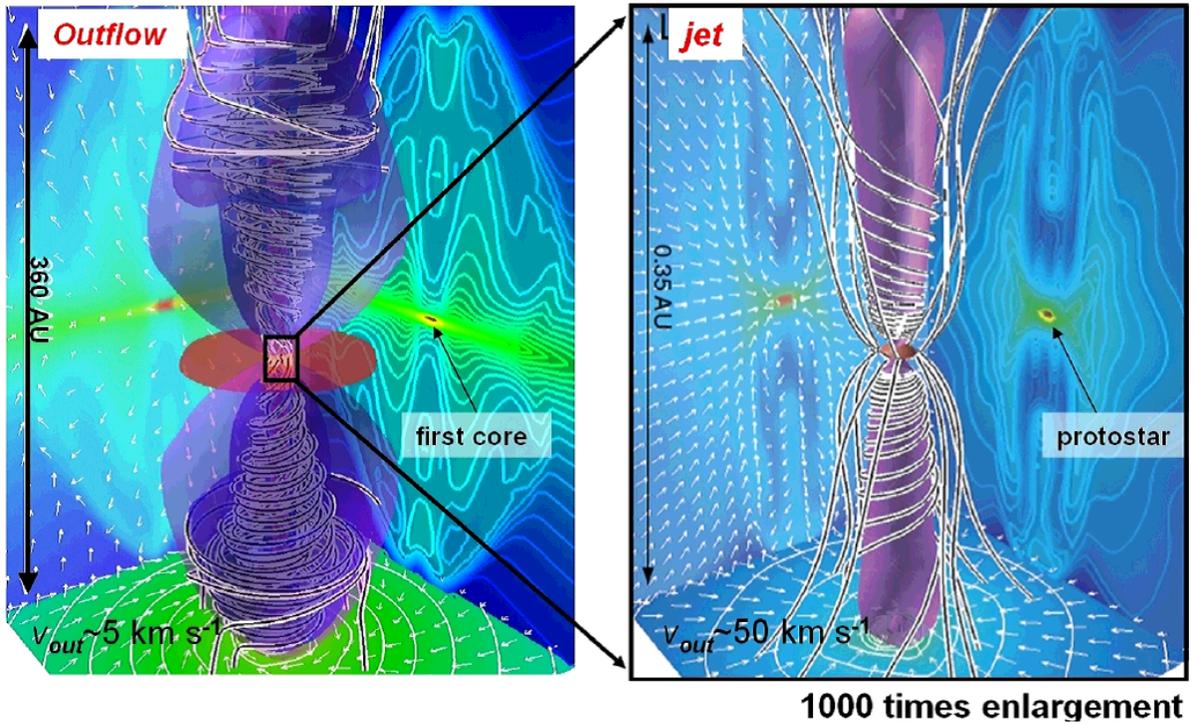} 
%%\vspace*{-0.3 cm}
\caption{
Low-velocity outflow driven by the first core (left panel) and high-velocity jet driven by the protostar (right panel).
The magnetic field lines are plotted by black-and-white streamlines.
Inside the purple surfaces, the flow is outflowing from the central object (the first core or protostar), and outside the purple surface (in the blue regions), the flow is inflowing to the central object.
}
\label{fig:2}
\end{center}
\end{figure}

Such two-component flows are naturally reproduced in recent star forming simulations, in which the star formation process is calculated from the prestellar stage until protostar formation (\citealt{tomisaka02}, \citealt{machida05a}, \citealt{machida06}, \citealt{machida08b}, \citealt{hennebelle08a}, \citealt{machida09a}, \citealt{banerjee06}, \citealt{duffin09}, \citealt{com10}, and \citealt{tomida10}).
As described in Figure~\ref{sec:outline}, two nested cores (the first core and protostar) appear in the star formation process, and each core can drive different types of flows.
The first core is formed in the low-density region ($n<10^{12}\cm$).
Thus, a relatively strong magnetic field surrounds the first core, because the first core does not experience Ohmic dissipation.
Note that Ohmic dissipation becomes effective within a range of $10^{12}\lesssim n \lesssim 10^{15}\cm$ (see, \S\ref{sec:mag}).
This strong magnetic field can drive a low-velocity outflow by the magneto-centrifugal mechanism (\citealt{blandford82}).
On the other hand, the protostar appears in the high-density region ($n>10^{21}\cm$).
Thus, an extremely weak magnetic field surrounds the protostar, because such a high-density region experiences Ohmic dissipation.
This weak field cannot drive outflow by the magneto-centrifugal mechanism.
Instead, the rotation of the protostar generates a strong toroidal field, and the magnetic pressure gradient force can drive a high-velocity flow.
As a result, the first core and the protostar drive two flows resulting in a low-velocity outflow surrounding a high-velocity jet as seen in Figure~\ref{fig:2}.

The different depths of the gravitational potential (or different Kepler velocities) cause the different outflow speeds.
The first core has a relatively shallow gravitational potential and drives a relatively slow outflow, while the protostar has a deeper gravitational potential and drives a high-velocity jet.
The different driving mechanisms for the outflow and jet cause the difference in the degrees of collimation.
The magneto-centrifugal mechanism drives the low-velocity outflow with wide opening angle, while the magnetic pressure gradient force along the rotation axis drives the high-velocity jet with good collimation (\citealt{machida08b}).
In summary, the different properties of the drivers (the first core and protostar) cause the difference in the properties of these two flows.

\section{Angular Momentum and Magnetic Flux Problems}
\label{sec:mag}
Molecular clouds have rotational energy equal to $\sim 2\%$ of their gravitational energy (e.g., \citealt{caselli02}), while they have magnetic energy comparable to their gravitational energy (e.g., \citealt{crutcher99}).
Conservation of the angular momentum and the magnetic flux in a collapsing cloud suggests that the rotation and the magnetic field in the cloud gradually increase as the cloud collapses, thus preventing further collapse and protostar formation. 
However, the rotation and magnetic field strength of the observed protostars indicate that neither the angular momentum nor the magnetic flux is conserved in collapsing clouds.
In general, these anomalies are referred to as the ``angular momentum problem'' and ``magnetic flux problem.'' 
The former problem is that the specific angular momentum of a molecular cloud is much larger than that of a protostar.
The latter problem refers to the fact that the magnetic flux of a molecular cloud is much larger than that of a protostar with equivalent mass.
These problems imply that there must be mechanisms for removing angular momentum and magnetic flux  from a cloud core.
In a collapsing cloud, these two problems are  related.
Namely, the angular momentum is removed by magnetic effects (i.e., magnetic braking, outflows, and jets), while the magnetic field is amplified by the shearing motion caused by cloud rotation.
Hence, the rotation and the magnetic field  cannot be treated independently while considering the angular momentum and magnetic flux problems.

Recently, the evolution of the angular momentum and magnetic flux in a collapsing cloud through protostar formation has been investigated (\citealt{machida07} and \citealt{duffin09}).
In the collapsing cloud, magnetic braking and protostellar outflow in the magnetically active regions ($n<10^{12}\cm$ and $n>10^{16}\cm$) remove the angular momentum.
By the time a protostar is formed, about 3$-$4 orders of magnitude of the initial angular momentum have been transferred by such magnetic effects.
Simulations suggest that the protostar at its formation has a rotation period of several days, which is comparable to the observations (\citealt{herbst07}).
In addition, when the number density exceeds $n\gtrsim10^{12}\cm$ in the collapsing cloud, the degree of ionization  becomes considerably low and Ohmic dissipation (and ambipolar diffusion) remove the magnetic flux.
Then, after the density exceeds $n\gtrsim10^{16}\cm$, thermal ionization of alkali metals reduces the resistivity and Ohmic dissipation becomes ineffective.
By the time a protostar forms, about 3$-$5 orders of magnitude of the initial magnetic flux has been removed within the range of $10^{12}\cm\lesssim n \lesssim 10^{15}\cm$.
Simulations suggest that the protostar at its formation has a sub-kilogauss magnetic field strength, which is also comparable to the observations (\citealt{bouvier2007}).
Thus, recent numerical simulations could resolve the angular momentum and magnetic flux problems in the early phase of the star formation (i.e., until the Class 0 phase).
However, to determine the rotation period and magnetic field strength of older protostars (Class I, II, and III phases) and main-sequence stars, we must further investigate their evolution.

\section{Circumstellar Disk Formation}
Stars form in molecular cloud cores that have nonzero angular momenta, thus, the appearance of a circumstellar disk is a natural consequence of star formation when angular momentum is conserved in the collapsing cloud.
In addition, observations have shown the existence of circumstellar disks around protostars.
Numerous observations indicate that the circumstellar disks around Class I and II protostars have sizes of $\sim10-1000\,$AU and masses of $\sim10^{-3}-0.1\msun$ (e.g., \citealt{natta00}).
%%However, such disks are images long after their formation. 
Because the formation sites of the circumstellar disk and protostar are embedded in a dense infalling envelope, it is difficult to directly observe newborn or very young circumstellar disks.
Thus, in general, we observe only the circumstellar disks long after their formation, i.e., around Class I or II protostars.
Observations also indicate that younger protostars have more massive circumstellar disks (e.g., \citealt{natta00} and \citealt{meyer07}).
Recently, \citet{enoch09} observed  massive disks with $M_{\rm disk} \sim 1 \msun$ around Class 0 sources, indicating that a massive disk can be present early in the main accretion phase.
However,  observations cannot determine the real sizes of circumstellar disks, or how and when they are formed.
Therefore, both theoretical approach and numerical simulations are necessary to investigate the formation and evolution of circumstellar disks.

In the collapsing cloud, before the protostar forms, the first core appears with a size of $\sim1-10$\,AU and mass of $\sim0.01-0.1\msun$.
Recent studies showed that the first core  directly evolves into a circumstellar disk after the protostar forms (\citealt{bate98}, \citealt{bate10}, \citealt{walch09}, \citealt{machida10}, \citealt{inutsuka10}).
The first core has a disk-like structure at its formation because the first core is supported not only by thermal pressure but also by rotation.
Thus, even after the protostar forms (or after the dissociation of molecular hydrogen begins), the first core does not disappear; instead it becomes a Keplerian rotating disk in the main accretion phase.
In summary, the protostar is formed inside the disk-like first core.
In other words, a massive circumstellar disk with size of $>1$\,AU already exists at the moment of the birth of the protostar.
In the main accretion phase, such a massive disk tends to show fragmentation, subsequently forming a binary companion or gas-giant planet.
Thus, recent numerical results support the concept that gravitational instability creates gas-giant planets.

\section{Summary}
\begin{figure}[b]
%%\vspace*{-0.2 cm}
\begin{center}
 \includegraphics[width=160mm]{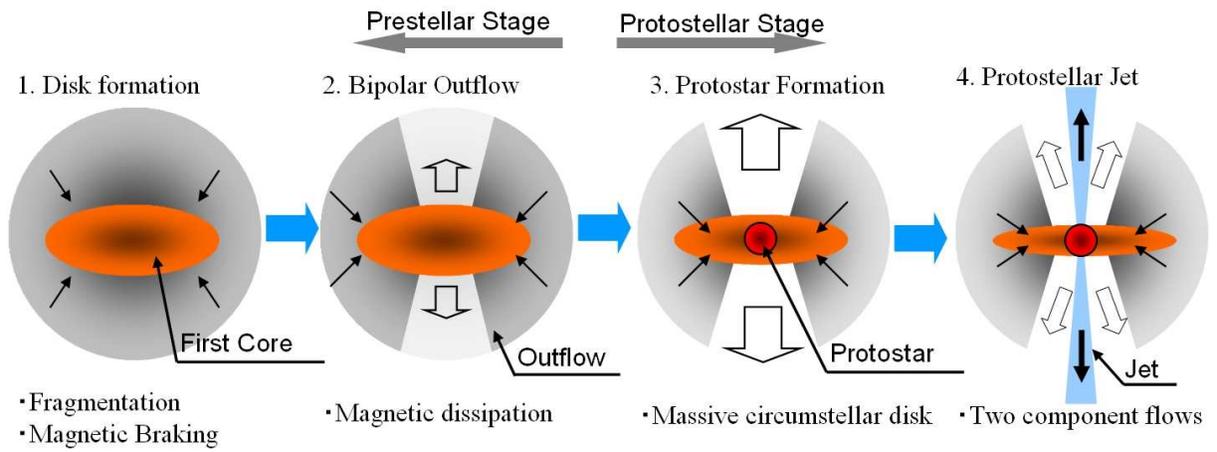} 
%%\vspace*{-0.2 cm}
\caption{Low-mass star formation scenario from the prestellar cloud core through protostar formation.}
\label{fig:3}
\end{center}
\end{figure}

Recent numerical simulations have changed the classical star formation scenario.
In Figure~\ref{fig:3}, I briefly summarize the new star formation scenario, suggested by these recent studies.
Gas collapse occurs around a small central part of the molecular cloud.
In the collapsing cloud core, the gas becomes adiabatic and the first adiabatic core appears prior to the protostar formation (stage 1).
In this stage, fragmentation frequently occurs to form binary or multiple star systems, because the gas collapse slows down and the perturbations that induce fragmentation can grow.
In addition, the first core can drive a low-velocity outflow with a wide opening angle, because the rotation timescale becomes shorter than the collapsing timescale and the magnetic field is amplified by the rotation.
Then, the amplified magnetic field drives outflow by the magneto-centrifugal mechanism.
This flow corresponds to the observed molecular outflow.
Also, in this stage, over 99\% (or 99.9\%) of the angular momentum of the central part of the cloud core is transferred by magnetic effects such as magnetic braking and outflows.
The first core increases its mass and density through  gas accretion.
Then, the magnetic field begins to dissipate through Ohmic dissipation when the central density exceeds $n>10^{12}\cm$ (stage 2).
Within the range of $10^{12}\cm \lesssim n \lesssim 10^{16}\cm$, the magnetic flux is largely removed from the collapsing cloud.
In this period, about 3$-$5 orders of magnitude of the initial magnetic flux of the collapsing cloud is removed.
The removal of angular momentum and magnetic flux  makes further collapse possible.
When the central density exceeds $n\sim10^{21}\cm$, the protostar appears (stage 3).
At the protostar formation epoch, the protostar is enclosed by the disk-like first core.
After the protostar formation, the first core  directly evolves into a circumstellar disk with Keplerian rotation.
Just after the protostar forms, a high-velocity jet with good collimation appears near the protostar (stage 4).
The magnetic field around the protostar is very weak because of Ohmic dissipation.
Thus, the high-velocity jet is driven by the magnetic pressure gradient force (or strong toroidal field) that is generated by the rotation of the protostar.
In addition, jet is well collimated, because it propagates along the rotation axis.
Moreover, the low-velocity outflow with a wide-opening angle continues to be driven by the circumstellar disk that originated from the first core.
Thus, a high-velocity jet is enclosed by a low-velocity outflow after protostar formation.
At the protostar formation epoch, the protostar and first core (or the circumstellar disk) have masses of $10^{-3}\msun$ and $0.01-0.1\msun$, respectively.
Thus, in the main accretion phase, the circumstellar disk is more massive than the protostar.
Such a massive disk tends to fragment due to gravitational instability, thus creating a binary companion or gas-giant planet in the circumstellar disk.

Recent numerical simulations have unveiled the protostar formation process starting from the prestellar core stage, while  protostellar evolution long after the protostar formation (Class I, II and III phases) remains veiled.
Further developments or long-term calculations starting from the prestellar core stage are necessary in order to understand the later phases of star formation.

\vspace{-0.3cm}

\clearpage
\vspace{-0.3cm}

\clearpage
\vspace{-0.3cm}

\end{document}